\documentclass[%
superscriptaddress,
 amsmath,amssymb,amsfonts,
pra,
twocolumn
]{revtex4-2}

\pdfoutput=1 
\usepackage{graphicx}% Include figure files
\usepackage{bm} % bold math
\usepackage{physics}
\usepackage{xcolor}
\usepackage{textcomp}
\usepackage{soul}
\usepackage[colorlinks=true, allcolors=black]{hyperref}

\newcommand{\G}{\mathbb{G}}
\newcommand{\M}{\mathbb{M}}
\newcommand{\Proj}{\mathbb{P}}
\newcommand{\ProjR}{\mathbb{P}_r}

\newcommand{\I}{\mathbb{I}}

\newcommand{\REVISION}[1]{#1}
\DeclareUnicodeCharacter{202F}{\REVISION{There is a 202F unicode character here}}

\begin{document}

\preprint{APS/123-QED}

\title{Physical limits on Raman scattering: \\ the critical role of pump and signal co-design
}

\author{Alessio~Amaolo}
\affiliation{Department of Chemistry, Princeton University, Princeton, New Jersey 08544, USA}
\email{alessioamaolo@princeton.edu}
\author{Pengning~Chao}
\affiliation{Department of Mathematics, Massachusetts Institute of Technology, Cambridge, Massachusetts 02139, USA}
\author{Thomas~J.~Maldonado}
\affiliation{Department of Electrical and Computer Engineering, Princeton University, Princeton, New Jersey 08544, USA}
\author{Sean~Molesky}
\affiliation{Department of Engineering Physics, Polytechnique Montréal, Montréal, Québec H3T 1J4, Canada}
\author{Alejandro~W.~Rodriguez}
\affiliation{Department of Electrical and Computer Engineering, Princeton University, Princeton, New Jersey 08544, USA}
\date{\today}

\begin{abstract}
We present a method for deriving limits on Raman scattering in structured media and exploit it to constrain the maximum Raman signal resulting from a planewave incident on either a single Raman molecule in the vicinity of a structured medium or a designable Raman medium. 
Results pertaining to metallic and dielectric structures illustrate the importance of accounting for the nonlinear interplay between pump and signal fields, showing that treating the pump-focusing and signal-extraction processes separately, as in prior work, leads to unrealistic enhancements. 
The formulation could readily find applications in further enhancing surface-enhanced Raman scattering (SERS) spectroscopy and Raman-assisted lasing.
\end{abstract}

\maketitle

\begin{figure}[t!]
\centering\includegraphics[width=\linewidth]{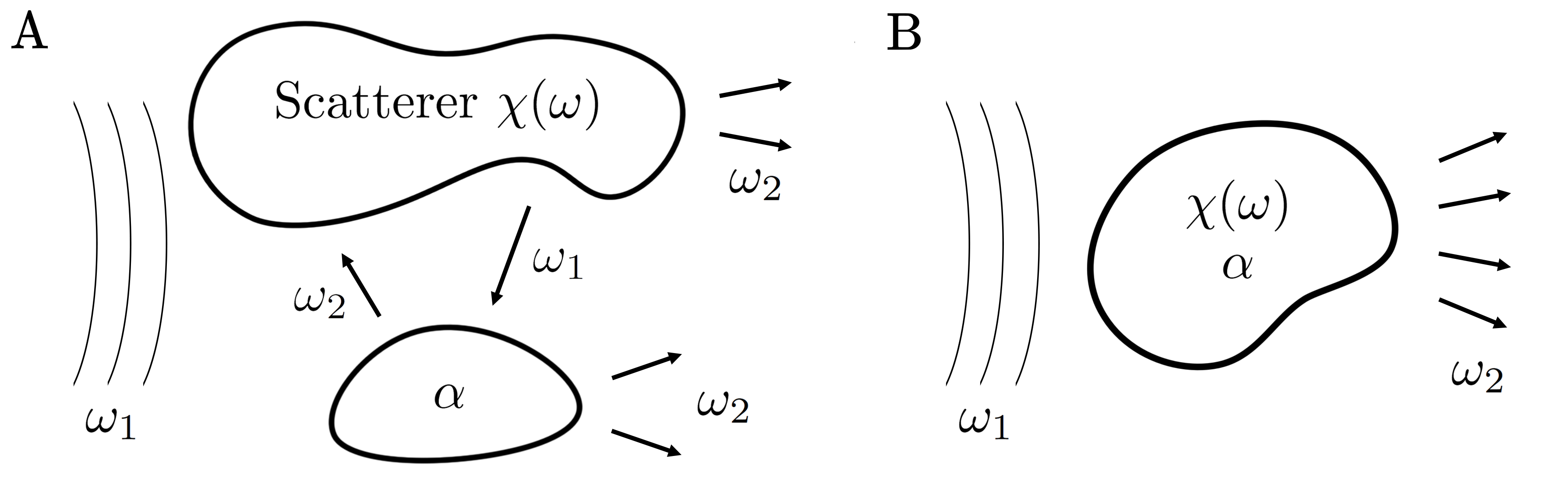}
    \caption{Schematic of the two configurations under consideration. (A) An electromagnetic wave is incident on an arbitrarily structured medium of \REVISION{linear} susceptibility \(\chi(\omega)\) in the vicinity of a known Raman-active region \(\alpha\). (B) An electromagnetic wave is incident on a structured scatterer of susceptibility \(\chi(\omega)\) and Raman polarizability \(\alpha\). In either case, we aim to maximize the Raman signal at \(\omega_2\).}
    \label{fig:schematic}
\end{figure}

\paragraph{Introduction.---} Raman scattering plays an important role in the development of spectroscopic methods~\cite{Le_Ru_Etchegoin_2009} and lasers~\cite{Rong_Jones_Liu_Cohen_Hak_Fang_Paniccia_2005}. While these processes are generally weak, Raman signals can be enhanced by nanostructuring~\cite{Han_Rodriguez_Haynes_Ozaki_Zhao_2022}, for example by focusing the source field or via the creation of optical resonances at the Raman frequency (or both)~\cite{Christiansen_Michon_Benzaouia_Sigmund_Johnson_2020}. Recent attempts to assess the limitations of structuring for enhancing Raman scattering~\cite{Michon_Benzaouia_Yao_Miller_Johnson_2019} have relied on linearized analyses that treat design constraints at the pump and signal frequencies independently. In this article, we present a quadratic optimization scheme to compute limits on Raman scattering that captures the inherent nonlinearity of this process and thus fully accounts for trade-offs in co-designing for the incident and Raman-scattered fields. 
We consider two configurations, depicted in Fig.~\ref{fig:schematic}: (A) light incident on a Raman scatterer in the vicinity of a structured  medium~\cite{Christiansen_Michon_Benzaouia_Sigmund_Johnson_2020, Long_Ju_Yang_Li_2022}, and (B) a designable Raman medium~\cite{Lu_Zhang_Chen_Shang_Huang_Liang_2022}. 
\REVISION{The theory presented is general and can treat widely applied experimental geometries, including improving surface- and tip-enhanced Raman spectroscopy (A), where the linear scatterer \(\chi(\omega)\) is a surface or a tip, or the efficiency or directivity of Raman gain media for lasing applications (B) via structuring.
In this paper, we study a representative example for each configuration.}
In the case of a \textit{Raman molecule} surrounded by a circular antenna, bounds on maximum achievable signal enhancement are shown to follow trends and values seen in optimized geometries. Conversely, limits that optimize focusing and scattering sub-problems independently of one another overestimate achievable performance by one to two orders of magnitude, demonstrating that modelling the fully coupled Raman problem is crucial. Compared to prior limits which also decouple and further relax this problem~\cite{Michon_Benzaouia_Yao_Miller_Johnson_2019}, the largest possible Raman signals are shown to be more than four orders of magnitude smaller.  In the case of a \textit{Raman medium}, previously unexplored, results reveal achievable performance coming within factors of unity of the bound when the pump and signal frequencies are close. 

The nascent ability to exploit convex optimization methods for deriving wave limits mirrors related advances in the area of structural optimization or ``inverse design"~\cite{Angeris_Vučković_Boyd_2019,Molesky_Chao_Mohajan_Reinhart_Chi_Rodriguez_2022,
Chao_Strekha_Kuate_Defo_Molesky_Rodriguez_2022, Kuang_Miller_2020,Gustafsson_Schab_Jelinek_Capek_2020}. 
In combination with structural optimization---an NP-hard problem that forbids efficient guarantees of optimal solutions---bounds offer a general-purpose (both geometry and mechanism agnostic) means of determining possible performance gaps. 
While originally confined to linear electromagnetic settings, with one recent exception~\cite{Mohajan_Chao_Jin_Molesky_Rodriguez_2023}, the present work demonstrates that quadratic optimization methods may be successfully adapted to study nonlinear wave problems, with potential applications to Raman lasers~\cite{Rong_Xu_Kuo_Sih_Cohen_Raday_Paniccia_2007}, frequency combs~\cite{Ideguchi_Holzner_Bernhardt_Guelachvili_Picque_Hänsch_2013}, and Raman amplification~\cite{Claps_Dimitropoulos_Raghunathan_Han_Jalali_2003}.

\paragraph{Formulation.---} 
To first order in the Raman response~\cite{Le_Ru_Etchegoin_2009}, within the rotating-wave approximation~\cite{Boyd_2003}, a source ``pump" field $\vb E_1(\vb r, \omega_1)$ oscillating at frequency $\omega_1$ incident on a Raman-active medium of polarizability \(\alpha \REVISION{\Proj_\alpha(\vb r)}\) generates a bound current \( \vb J_2 \equiv -i \omega_2 \alpha \Proj_\alpha(\vb r) \vb E_1 \) oscillating at the Raman ``signal" frequency $\omega_2$, \REVISION{where \(\alpha\) is a constant Raman polarizability and \(\Proj_\alpha\) is a projector encoding the spatial distribution of the Raman medium.}
Maxwell's equations (ignoring down-conversion) thus take the form
\begin{subequations} \begin{align} 
    \M_1 \vb E_1(\vb r, \omega_1) &= i \omega_1 \vb{J}_\textrm{vac}(\vb r, \omega_1)  \label{eq:eom1} \\
    \M_2 \vb E_2(\vb r, \omega_2) &= i \omega_2 \vb J_2 (\vb r, \omega_2) = \omega_2^2 \alpha \Proj_\alpha \vb E_1, \label{eq:eom2}
\end{align} \end{subequations}
where \(\M_j = \nabla \times \nabla \times - k_j^2 \REVISION{\left( 1+ \chi(\omega_j)\right) \Proj_\chi} \) for \(j=\{1,2\}\) are the linear Maxwell operators for the pump and signal wavelengths, \(\vb E_1 \) is the field sourced by the initial current $\vb{J}_\textrm{vac}$, \( \vb E_2 \) is the scattered Raman field, \(k_j = \omega_j/c\), \REVISION{\(\chi(\omega_j)\Proj_\chi\) is the susceptibility profile at \(\omega_j\) with constant susceptibility \(\chi(\omega_j)\) and projection operator \(\Proj_\chi\)}, \(\epsilon_0 = 1\), and we set the permeability \(\mu=1\). 
\REVISION{Intuitively, the pump field polarizes both the linear material and the Raman-active material. The former scatters the electric field following Maxwell's equations, while the latter produces a new current at the Raman frequency, producing a new electric field which is in turn scattered by the same linear material with susceptibility \(\chi({\omega_2})\).}
These equations are non-linearly coupled in two important ways. First, the field determined by~\eqref{eq:eom1} acts as a source in~\eqref{eq:eom2}. 
Second, the spatial profile of the susceptibility is the same at both frequencies.

A key figure of merit in determining possible Raman enhancement from structuring is the scattered power at the Raman frequency,
\begin{equation} \label{eq:opt_problem_full}
\begin{aligned}
    \underset{\Proj_\chi}{\text{max}} \quad & -\dfrac12 \Re \int \vb J_2^* \cdot \vb E_2 - \dfrac{\omega_2}{2} \int \vb E_2^* \Im \REVISION{\left( 1+ \chi(\omega_j)\right) \Proj_\chi} \vb E_2 \\
    \text{s.t.} \quad & \M_1 \vb{E}_1 = i\omega_1 \vb J_\text{vac} \\
     \quad & \M_2 \vb{E}_2 = \omega_2^2 \alpha \Proj_\alpha \vb E_1. %\\
     % \quad & \V(\vb r, \omega_j) = \left(1 + \chi(\omega_j)\right)
     % \Proj(\vb r)
\end{aligned} \end{equation} 
\(\Proj_\chi\) encodes the spatial susceptibility profile of the linear medium, and thus couples the two scattering problems. 
In~\eqref{eq:opt_problem_full}, the first term and second term of the objective function correspond to extracted (LDOS) and absorbed power, respectively.
% If the design material is itself Raman-active, we simply take \(\alpha(\vb r) = \alpha \Proj(\vb r)\) for a constant polarizability tensor \(\alpha\).
Intuitively, this quantity is often understood and optimized by relaxing the problem into two separate linear scattering processes. First, a field-focusing scheme in which the amplitude of the Raman source, proportional to the induced polarization \(\alpha \Proj_\alpha \vb E_1 \) from the incident source, can be maximized:
\begin{equation} \label{eq:focusing} \begin{aligned}
    \underset{\Proj_\chi}{\text{max}} \quad & \int \abs{\alpha \Proj_\alpha \overline{\vb E}_1}^2 \\
    % \overline{\vb E}_1^* \Proj_\alpha^\dagger \alpha^\dagger \cdot \alpha \Proj_\alpha \overline{\vb E}_1 \\ 
    \text{s.t.} \quad & \M_1  \overline{\vb E}_1 = i \omega_1 \vb J_{\textrm{vac}}.
\end{aligned} \end{equation}
Second, the scattered power at the Raman signal may be separately enhanced by solving 
\begin{equation} \label{eq:scat} \begin{aligned}
     \underset{\Proj_\chi}{\text{max}} \quad & -\dfrac12 \Re \int \left(\overline{\vb J}_2\right)^* \cdot \vb E_2 \\ & - \dfrac{\omega_2}{2} \int \vb E_2^* \REVISION{\Im \left( 1+ \chi(\omega_j)\right) \Proj_\chi} \vb E_2 \\
     \text{s.t.} \quad & \M_2 \vb{E}_2 = i\omega_2 \overline{\vb J}_2,
\end{aligned} \end{equation}
where the amplitude of the Raman current source \(\overline{\vb J}_2 = -i \omega_2 \alpha \Proj_\alpha \overline{\vb E}_1\) is obtained upon solving the field-focusing optimization problem. 
\REVISION{Intuitively, we are finding the maximum possible pump field at \(\Proj_\alpha\) subject to structuring. Then, we are utilizing the amplitude of this field to source the Raman current, and optimizing an independent photonic structure to maximize the Raman scattered power.}
Note that such a relaxation requires prior knowledge of \REVISION{\(\Proj_\alpha\)} and therefore cannot be used to bound problems where the design material is Raman-active. 
While solving~\eqref{eq:focusing} and~\eqref{eq:scat} separately offers conceptual and computational simplicity, leveraged by existing works~\cite{Michon_Benzaouia_Yao_Miller_Johnson_2019} to bound extracted power, \REVISION{the freedom to optimize \(\Proj_\chi\) independently for both problems} means that solutions can overestimate possible enhancements. 
%the explicit lack of a unique structure, \(\V(\vb r, \omega_j) = \chi(\omega_j) \Proj (\vb r) \) for all \(j \in \{1,2\}\), means that solutions can overestimate possible enhancements. 
The scattering processes of~\eqref{eq:focusing} and~\eqref{eq:scat} are not related by reciprocity, meaning the different structural considerations required at each wavelength would suggest higher performance than actually possible if treating these two processes separately. 
Maintaining the full coupling and nonlinear interplay between pump and signal fields, as achieved in this work, is therefore critically important for assessing co-design trade-offs.

\begin{figure*}[t]
    \centering
\includegraphics[width=\textwidth]{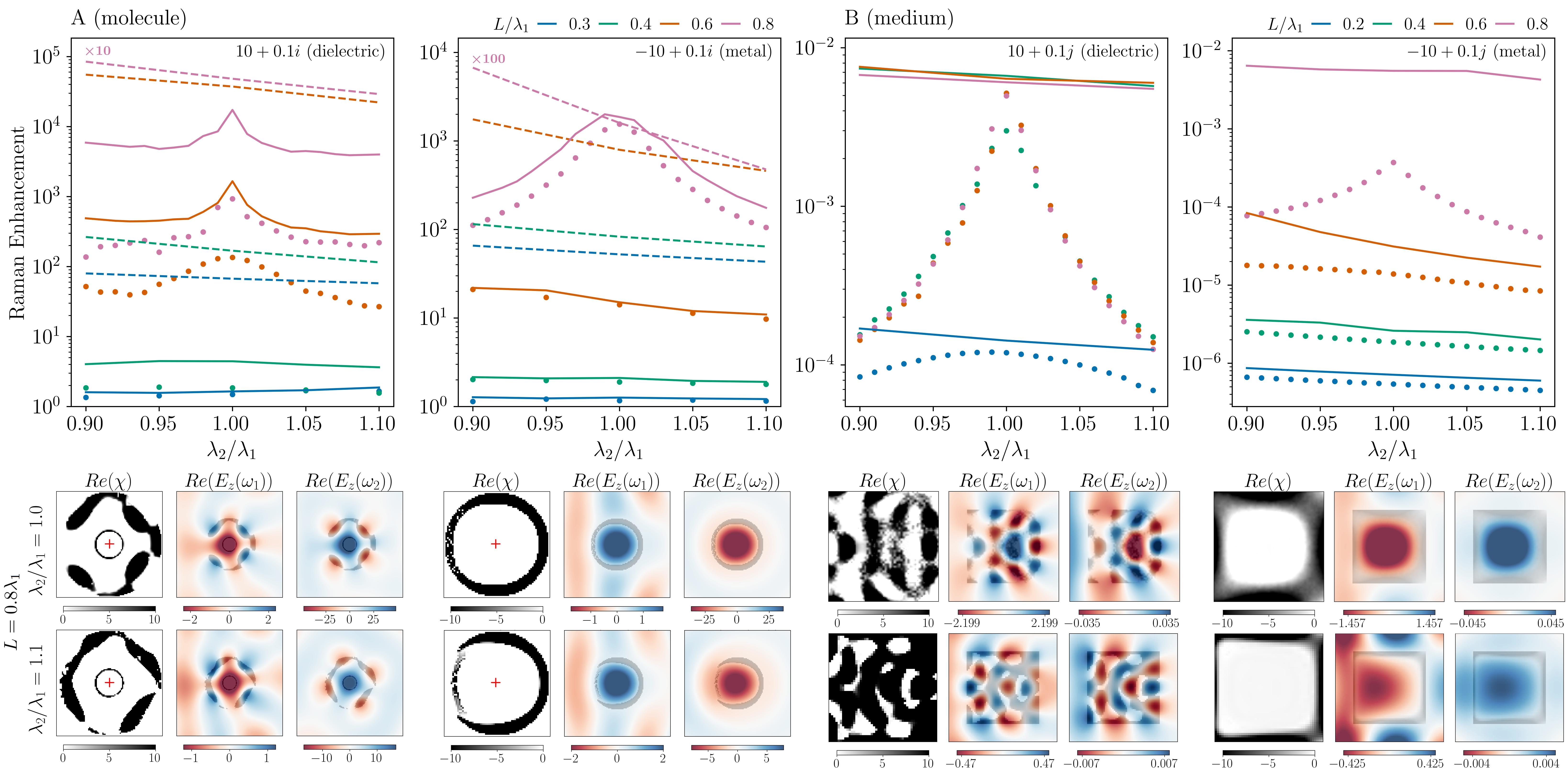}
    \caption{Limits (lines) and associated inverse designs (dots) pertaining to the maximum achievable Raman enhancement from an incident planewave on either (A) a Raman molecule \REVISION{of Raman polarizability \(\alpha = 1\)} \REVISION{(marked by a red plus)} surrounded by a structured medium (an antenna) contained within a \REVISION{2D} circular design region of inner diameter \(d/\lambda_1 = 0.2\) and \REVISION{variable} outer diameter \REVISION{\(L/\lambda_1 = \{0.3,0.4,0.6,0.8\}\)} or (B) a Raman-active medium \REVISION{of Raman polarizability \(\alpha = 0.01\)} contained within a square of \REVISION{variable} side-length \REVISION{\(L/\lambda_1 = \{0.3,0.4,0.6,0.8\}\)}.
    \REVISION{Solid lines show bounds via a fully coupled treatment of the nonlinear processes, while dashed lines show bounds obtained via linearization of the Raman molecule problem (see SM~\cite{supp}), demonstrating the importance of modeling coupling between the field-focusing and scattering processes.}
    The Raman enhancement is defined as the ratio of net scattered power \(f_\text{scat}\) in the presence and absence of the medium (A), or \(f_\text{scat}\) over the incident power (B). Both dielectric (\(\chi_{1,2} = 10+0.1i\)) and metallic \((\chi_{1,2}=-10+0.1i\)) media are considered.
    \REVISION{Representative inverse designs for \(\lambda_2 / \lambda_1 = \{1.0, 1.1\}\) and \(L/\lambda_1 = 0.8\) are shown with their respective field profiles for all cases, highlighting the importance of engineering field profiles at both pump and Raman frequencies.}}
    \label{fig:combined}
\end{figure*}

We first consider the case of a \textit{\REVISION{known Raman scatterer}}, illustrated schematically in Fig.~\ref{fig:schematic}(A).
\REVISION{In particular, for a point molecule, \(\Proj_\alpha = \delta(\vb r - \vb r')\).} 
We seek to maximize objectives of the fields with respect to the susceptibility distribution of an adjacent linear structure \(\Proj_\chi\). 
Defining $\G_j$ as the operator form of the vacuum Green's function, $(\nabla \times \nabla \times - k_j^2 )\G_j = k_j^2 \I$, we can decompose \(\vb E_1\) into its incident and scattered components: \( \vb E_1 = \vb E_{i,1} + \vb E_{s,1} = \vb E_{i,1} + \frac{iZ}{k_1} \G_1 \vb P_1 \) where \(Z\) is the vacuum impedance and \(\vb P_1\) the polarization field, which is zero in vacuum regions. Correspondingly, the current at \(\omega_2\) is given by
\begin{equation} \label{eq:J2}
    \vb J_2 = -i \omega_2 \alpha \dfrac{iZ}{k_1} \ProjR \left( \vb S_1 + \G_1 \vb P_1 \right)
\end{equation}
where \(\vb S_1 = \frac{k_1}{iZ} \vb E_{i,1} \) is defined for convenience. 
Note that the global phase factor in~\eqref{eq:J2} is irrelevant for the objective in~\eqref{eq:scat} and that we assume all Raman sources in \REVISION{$\Proj_\alpha$} are emitting in phase; this assumption trivially holds for a single Raman molecule. 
\REVISION{This will lead to higher objective values in~\eqref{eq:opt_problem_full} than if fully modeling random phases, which leads to a significantly more difficult inverse design~\cite{yao_johnson_distributed_raman_2023} problem; bounding such processes is the subject of future work~\cite{chao_distributed_raman_2024}. Regardless, bounds on~\eqref{eq:opt_problem_full} assuming in-phase emission are bounds on the corresponding random-phase problem.}
The total field at \( \omega_2 \) can then be expressed in terms of the polarization field \(\vb P_2 (\vb r, \omega_2) \) via
\begin{equation}
    \vb E_2 = \dfrac{iZ}{k_2} \G_2 \left( \vb J_2 + \vb P_2 \right).
\end{equation}
where again we neglect higher-order Raman effects such as down-conversion and work within the undepleted-pump approximation~\cite{Boyd_2003}. 
Following the same procedure introduced in recent works~\cite{Chao_Strekha_Kuate_Defo_Molesky_Rodriguez_2022}, one can write the optimization problem over Raman fields as a Quadratically Constrained Quadratic Program (QCQP) over the polarization fields \(\vb P_1, \vb P_2 \), with quadratic (energy-conservation) constraints derived by taking known inner products of operator constraints (see Suppplementary Material~\cite{supp}). The resulting QCQP takes the form 
\begin{equation} \label{eq:particle_opt} \begin{aligned}
    \underset{\{\vb P_1, \vb P_2\}}{\text{max}} \quad & f(\vb P_1, \vb P_2) \\
    \text{s.t.} \quad 
    & \int_{\Omega_k} \vb S_i^\dagger \cdot \vb P_j d\vb r = \int_{\Omega_k} \vb P_i^\dagger \left( \chi_i^{-\dagger} - \G_{i}^\dagger \right) \cdot \vb P_j  d\vb r, \\
    & \dfrac{\Im \chi_2}{\abs{\chi_2}^2} \int_{\Omega} \vb P_2^\dagger \cdot \vb P_2 \leq \int_{\Omega} \vb J_2^\dagger \G_{2}^\dagger \cdot \G_{2} \vb J_2 d\vb r
\end{aligned} \end{equation}
where \(i,j \in \{1,2\} \), \(\chi_{i}\) is the bulk susceptibility of the linear medium at \(\omega_i\), \(f\) is the quadratic objective function, and \(\vb S_2 = \G_{2} \vb J_2 \) may be interpreted as the Raman-induced field incident on the linear medium.
A full derivation of these constraints, including expanding all instances of \(\vb J_2\) in terms of the optimization variables \(\vb P_1, \vb P_2\), is presented in the SM~\cite{supp}.
The imposed constraints hold for all and any sub-regions \(\Omega_k \subset \Omega\) of the total design region, \REVISION{and can be interpreted as enforcing Poynting's theorem in all regions.} 
As described in the SM~\cite{supp}, the last constraint, which follows from passivity, is required only for numerical reasons. 
The solution of the Lagrange dual to~\eqref{eq:particle_opt} for a given \(\Proj_\alpha\) is a geometry-independent bound on~\eqref{eq:opt_problem_full}.
Detailed calculations of the corresponding Lagrange dual, gradients, and proof of the existence of a bound are also shown in the SM. 

For the \textit{Raman medium} case illustrated in Fig.~\ref{fig:schematic}(B), the structured medium itself is the Raman scatterer, so that \REVISION{\(\Proj_\alpha = \Proj_\chi\)}. Here, we relax \(\vb J_2\) to an optimization degree of freedom and enforce inner products over~\eqref{eq:J2} as additional constraints. 
Maintaining the same definitions as above, the resulting QCQP takes the form 
\begin{equation}
    \begin{aligned}
        \underset{ \{ \vb P_1, \vb P_2, \vb J_2 \} }{\text{max}} \quad & f(\vb P_1, \vb P_2, \vb J_2)      \\
    \text{s.t.} \quad & \int_{\Omega_k} \vb S_i^\dagger \cdot \vb P_j d\vb r = \int_{\Omega_k} \vb P_i^\dagger \left( \chi_i^{-\dagger} - \G_{i}^\dagger \right) \cdot \vb P_j  d\vb r, \\
                    & \int_{\Omega_k} \vb S_i^\dagger \cdot \vb J_2 d\vb r = \int_{\Omega_j} \vb P_i^\dagger \left( \chi_i^{-\dagger} - \G_{i}^\dagger \right) \cdot \vb J_2  d\vb r, \\  
                    & \int_{\Omega_j} \vb J_2^\dagger \cdot \vb J_2 d\vb r = \abs{\omega_2}^2 \int_{\Omega_j} \abs{\alpha \Proj_\alpha \vb E_1}^2 d \vb r
                    %\abs{\dfrac{\omega_2}{\omega_1 \chi_1}}^2 \int_{\Omega_j} \abs{\alpha \Proj_\alpha \vb P_1}^2 d\vb r
                    %\vb P_1^\dagger \alpha^\dagger \cdot \alpha \vb P_1  d\vb r,
    \end{aligned}
\end{equation}
Note that the last constraint is phase-agnostic and that by promoting \(\vb J_2\) to an optimization degree of freedom, we allow variations in its phase profile.
A bound on this problem is therefore also a bound on a medium containing a uniform distribution of fluctuating, spatially uncorrelated (out-of-phase) Raman scatterers~\cite{Yao_Verdugo_Everitt_Christiansen_Johnson_2023}. Calculation of the Lagrange dual is presented in the SM~\cite{supp}.

\paragraph{Applications.---} 
In what follows, we exploit the framework above to obtain upper bounds on the maximum allowed  scattered power \(f_\text{scat}(\vb P_1, \vb P_2) = -\frac12 \Re \int_{\Omega} \vb J_2^* \cdot \vb E_2 d\vb r - \int_{\Omega} \vb P_2^* \frac{\Im \chi_2}{\abs{\chi_2}^2}\frac{Z}{2k_2} \vb P_2 d\vb r \) achievable in two canonical typical configurations. 
\(f_\text{scat}\) can be written purely in terms of coupled \(\vb P_1, \vb P_2 \) fields (see SM). 
To assess the importance of co-designing for both pump and signal processes, we also compare results of bounding the full problem of \eqref{eq:opt_problem_full} for a Raman molecule against those made possible via the linearized optimization problems \eqref{eq:focusing} and \eqref{eq:scat}.

\textit{Raman molecule:} we first consider the maximum achievable scattered power \(f_\text{scat}\) that may arise \REVISION{in two dimensions} from a TM (electric field out of plane) planewave of wavelength \(\lambda_1\) incident on a Raman particle of polarizability \( \alpha \delta(\vb r - \vb r_0) \) \REVISION{with \(\alpha = 1\)} at the center \(\vb r_0\) of a
circular design region of inner diameter \(d/\lambda_1=0.2\) and outer diameter \(L/\lambda_1 \in \{0.3, 0.4, 0.6, 0.8 \}\). 
The Raman enhancement factor is defined as \(f_\text{scat}(\vb P_1, \vb P _2) / f_\text{scat}(\vb 0)\), where \(f_\text{scat}(\vb 0)\) represents the extracted power in the absence of the linear scatterer \REVISION{(\(\chi(\omega_j) = \vb 0 \))}.
The circular design region consists of either a dielectric (\( \chi_1 = \chi_2 = 10+0.1i\)) or metallic (\(\chi_1 = \chi_2 = -10+0.1i\)) medium. 
As shown in Fig.~\ref{fig:combined}(A), bounds (solid lines) follow trends seen in topology-optimized designs (dots) over a broad range of signal wavelengths \(\lambda_2/\lambda_1 \in [0.9,1.1]\), peaking at the resonant condition \(\lambda_2/\lambda_1 = 1 \).
As seen, the performance of dielectric structures is generally found to be within an order of magnitude of the bound for \(L \leq 0.6\lambda_1\), becoming worse with increasing system size. Bounds on metals not only predict trends, anticipating the minimum design size needed to reach the resonant regime (the peak near \(\lambda_2/\lambda_1 \sim 1\)), but are also within factors of unity of inverse designs.
Note that when adapted to bound extracted power, our performance metrics improve upon existing state-of-the-art~\cite{Michon_Benzaouia_Yao_Miller_Johnson_2019} by more than four orders of magnitude (see SM~\cite{supp}). 

\begin{figure}
    \centering
    \includegraphics[width=0.45\textwidth]{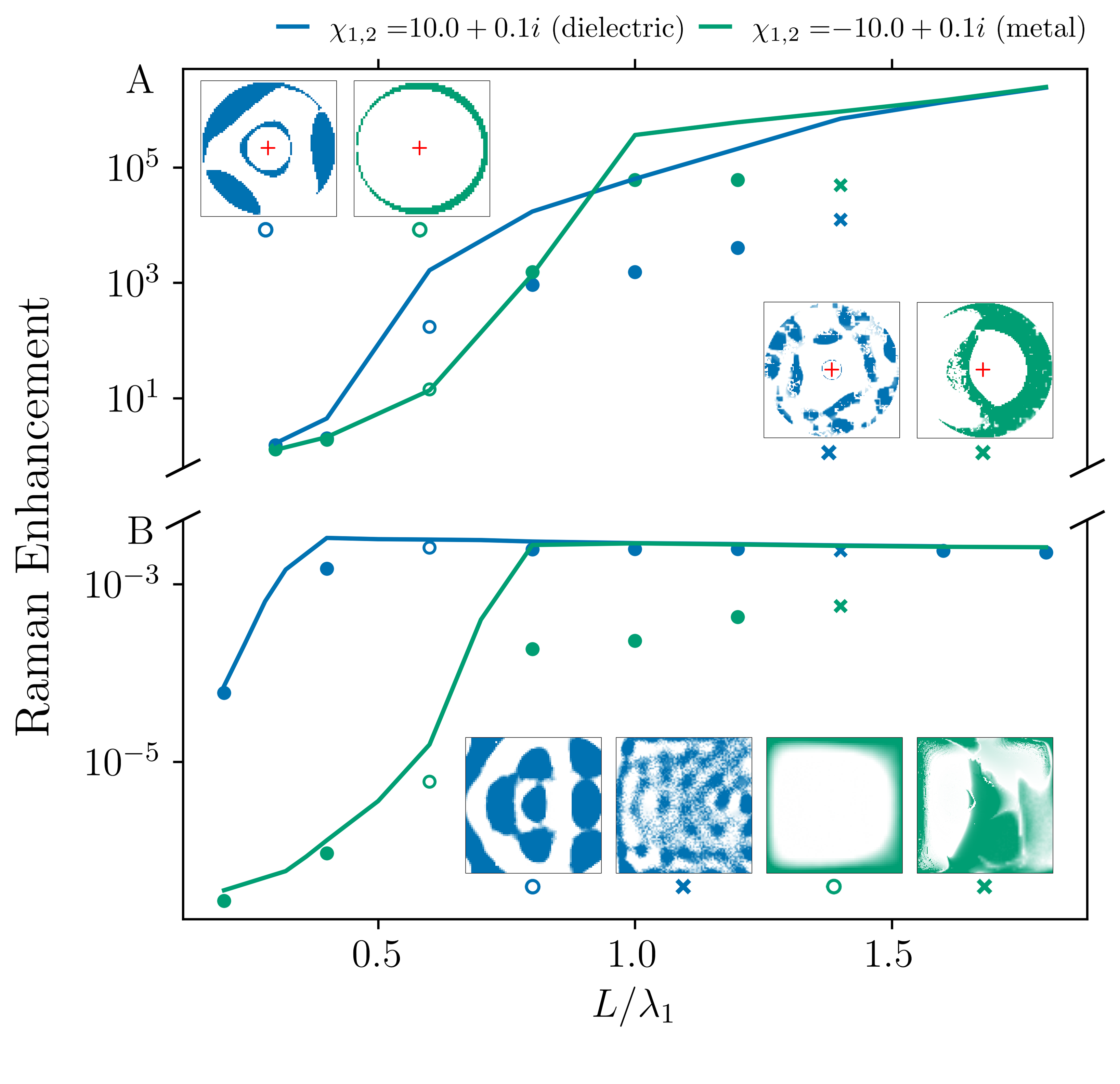}
    \caption{
    Limits (lines) and associated inverse designs (dots) pertaining to maximum \(f_\text{scat}\) as a function of device size $L/\lambda$ for equivalent problem formulations as in Fig.~\ref{fig:combined}, with \(\lambda \equiv \lambda_1 = \lambda_2\). \REVISION{(A) and (B) panels show results pertaining to either \textit{Raman molecules} (\(\alpha = 1\)) or \textit{Raman media} (\(\alpha = 0.01\)), respectively.}}
    \label{fig:scaling}
\end{figure}

Field profiles reveal resonances exhibiting high field concentrations near the Raman molecule and large coupling to the incident planewave at \(\lambda_1\), illustrating the relative difficulty and importance of co-designing two resonances at different frequencies. 
To quantify the importance of structures designed for simultaneous operation at both wavelengths as dictated by the fully coupled nonlinear problem of~\eqref{eq:opt_problem_full}, Fig.~\ref{fig:combined}(A) also shows limits (dashed lines) obtained by solving the ``linearized", independent problems of~\eqref{eq:scat} and~\eqref{eq:focusing}, as outlined in the SM~\cite{supp}. 
In particular, the relaxation overestimates achievable signals by up to four orders of magnitude, failing to capture the design trade-offs and performance costs incurred by the need to simultaneously enhance both processes.
This overestimation occurs even when \(\lambda_1=\lambda_2\) owing to the different design criteria necessary to achieve field focusing or scattering.
Finally, Fig.~\ref{fig:scaling} shows the scaling behavior of the maximum achievable scattered power \(f_{scat}\) with respect to system size \(L/\lambda_1\) in the resonant case \(\lambda_2=\lambda_1\). 
For large enough systems, metals and dielectrics are found to converge to same maximum performance, suggesting that given sufficient structural freedom, the maximum Raman response is not limited by material choice. While dielectrics provably outperform metals in the subwavelength regime, this is expected to reverse for TE (in-plane electric) fields.

\textit{Raman medium:} next, we consider the maximum achievable scattered power \(f_\text{scat}\) that may arise from a TM planewave of wavelength \(\lambda_1\) incident on a Raman-active square design region of either dielectric (\(\chi_1=\chi_2=10+0.1i\)) or metal (\(\chi_1=\chi_2=-10+0.1i\)) susceptibilities, Raman polarizability \(\alpha = 0.01 \), and of side length \(L/\lambda_1 \in \{0.3, 0.4, 0.6, 0.8 \}\), over a range of \( \lambda_1/\lambda_2 \in [0.9,1.1] \). 
The Raman enhancement factor is defined as the ratio of scattered power at \(\omega_2\) to incident power at \(\omega_1\).
As shown in Fig.~\ref{fig:combined}(B), bounds come within an order of magnitude of achievable performance when no resonant peak is present. Limits pertaining to dielectric structures accurately predict performance of optimized devices when the Raman shift is negligible (\(\lambda_1 \sim \lambda_2\)), while failing to capture non-resonant behavior. Similar behavior is observed for metals, except that the performance gap grows with increasing system sizes to over an order of magnitude at the largest \(L=0.8\lambda\) sizes explored. 
As in the case of Raman particles, field profiles show metallic designs supporting resonances at both the pump and Raman wavelengths and demonstrate the non-reciprocity of the focusing and scattering problems.
Unlike the previous case, structural information of the Raman scatterer is not known a priori so the bounds cannot be compared to their decoupled counterparts. 
Finally, as shown in Fig.~\ref{fig:scaling}, the scaling behavior of the bounds on resonant (\(\lambda_1=\lambda_2\)) Raman enhancement with \(L/\lambda_1\) come within factors of unity of achievable performance for dielectrics for all device sizes and within an order of magnitude for metals when the device is sufficiently small (\(L/\lambda_1 \leq 0.3 \)) or large (\(L/\lambda_1 \geq 0.6\)). The shrinking gap between performance and bound for large device sizes is attributable to the saturation of the limits to a response not limited by material choice.

\paragraph{Concluding remarks.---}  
While the typical approach of separately treating field-focusing and signal-extraction processes offers conceptual and computational simplifications, the presented comparisons of bounds that either incorporate or ignore structural coupling between the pump and Raman fields reveals the necessity of modelling the full non-linear problem.
Such an analysis was made possible by generalizing a recent quadratic optimization method previously restricted to linear electromagnetic problems~\cite{Chao_Strekha_Kuate_Defo_Molesky_Rodriguez_2022}. By taking into account full structural coupling in the objective, these results not only greatly improve on existing bounds~\cite{Michon_Benzaouia_Yao_Miller_Johnson_2019} but also pave the way for studying designable Raman-active structures.

\begin{acknowledgments}
\textit{Acknowledgements.---} We would like to thank Jewel Mohajan and Steven G. Johnson for useful discussions. 

\textit{Funding.---} We acknowledge the support by a Princeton SEAS Innovation Grant and by the Cornell Center for Materials Research (MRSEC). 
SM also acknowledges financial support from IVADO (Institut de valorisation des données, Québec).
The simulations presented in this article were performed on computational resources managed and supported by Princeton Research Computing, a consortium of groups including the Princeton Institute for Computational Science and Engineering (PICSciE) and the Office of Information Technology's High Performance Computing Center and Visualization Laboratory at Princeton University.
\end{acknowledgments}

\bibliography{citations}
\end{document}

\title{Supplemental Material for ``Physical limits on Raman scattering: the critical role of pump and signal co-design"}

\author{Alessio~Amaolo, Pengning~Chao, Thomas~J.~Maldonado, Sean~Molesky, Alejandro~W.~Rodriguez}
\date{\today}

\maketitle
\section{Duality Relaxation of Quadratically Constrained Quadratic Programs}
In this section, we show bounds can be derived via duality relaxation for a Quadratically Constrained Quadratic Program (QCQP). 
This discussion is not specific to the Raman problem and has been described in detail in Refs.~\cite{chao_physical_2022, Molesky_Chao_Jin_Rodriguez_2020}, but is partially reproduced here for clarity. 
\\~\\
\textbf{Calculation of Lagrange dual.} To simplify algebra and complicated integrals, these sections will make use of braket notation to mean conjugated inner products. 
We first write the Lagrangian 
\begin{equation}
\label{eq:lagrangian}
    \Lag (P_\text{opt},S) = 
    \begin{bmatrix}
    \bra{P_\text{opt}} & \bra{S} 
    \end{bmatrix}
    \begin{bmatrix}
    -Z^{TT}(\lambda) & Z^{TS}(\lambda) \\
    Z^{ST}(\lambda) & Z^{SS}(\lambda)  
    \end{bmatrix} 
    \begin{bmatrix}
    \ket{P_\text{opt}} \\
    \ket{S} 
    \end{bmatrix}.
\end{equation}
The Lagrange dual is defined as \(\mathcal{G}(\lambda) = \max_{P_\text{opt}} \Lag(\ket{P_\text{opt}}, \lambda) \) and can be written in terms of \(\ket{P^*_\text{opt}} = Z^{TT-1} Z^{TS}\ket{S}\) and for positive definite \(Z^{TT}\) as
\begin{equation} \begin{aligned}
    \mathcal{G}(\lambda) &= \bra{S}Z^{ST} Z^{TT-1} Z^{TS} + Z^{SS} \ket{S}.
    \label{eq:dual_ZTT}
\end{aligned} \end{equation}
Any feasible \(\mathcal{G}(\lambda)\) is a bound on the primal problem~\cite{Boyd_Vandenberghe_2004}. 
We may optimize the dual function by calculating its derivatives with respect to \(\lambda\),
\begin{equation}
    \dfrac{\partial\mathcal{G}}{\partial \lambda} = 2 \Re \left( \bra{P_\text{opt}^*} \dfrac{\partial Z^{TS}}{\partial \lambda} \ket{S} \right) - \bra{P_\text{opt}^*} \dfrac{\partial Z^{TT}}{\partial \lambda} \ket{P_\text{opt}^*} + \bra{S} \dfrac{\partial Z^{SS}}{\partial \lambda} \ket{S},
\end{equation}
and using gradient based optimization methods. 
By finding \(\min_\lambda \mathcal{G}\), the tightest possible limit on the given optimization problem can be found. 
In Eq.~\eqref{eq:lagrangian}, \(Z^{TT}, Z^{TS}, Z^{SS}\) contain both objective and constraint information. 
The objective for the Raman problem will be derived in the next section. 
\\~\\
\textbf{Derivation of QCQP Constraints.}
It is possible to derive the energy conservation constraints presented via operator constraints (which, like Maxwell's equations are dependent on the spatial profile of permittivity) relaxed to structure-agnostic scalar constraints~\cite{chao_physical_2022}. 
Although these constraints encode less information than Maxwell's equations in full generality, they allow the relaxed problem to be formulated as a QCQP.
In this section, we show how these constraints can be directly derived from energy conservation as expressed by Poynting's theorem for time-harmonic complex fields~\cite{tsang_scattering_2004}:
\begin{equation}
    \int_{\partial V} \dd\vb{\sigma} \cdot (\vb{E}\cross\vb{H}^*) = i\omega \int_V (\vb{H}^*\cdot\mu\cdot\vb{H} -  \vb{E}\cdot\epsilon^*\cdot\vb{E}^*) \,\dd V
    - \int_V \vb{E}\cdot\vb{J}^* \,\dd V. 
    \label{eq:Poynting_cplx}
\end{equation} 
For simplicity, we will assume a non-magnetic material $\mu=\mu_0=1$ and scalar isotropic permittivity and susceptibility $\epsilon = 1 + \chi$. 

Consider a scattering theory picture where a free current source $\vb{J}_v$ generates the fields $\vb{E}_v$, $\vb{H}_v$ in vacuum and $\vb{E}_t$, $\vb{H}_t$ in the presence of a structure with material distribution $\epsilon(\vb{r}) = 1 + \chi \I_s(\vb{r})$ where $\I_s(\vb{r})$ is an indicator function. There is an induced polarization current $\vb{J}_s$ in the material which produces scattered fields $\vb{E}_s$ and $\vb{H}_s$ that combine with the vacuum fields to give the total field: $\vb{E}_t = \vb{E}_v+\vb{E}_s$, $\vb{H}_t = \vb{H}_v + \vb{H}_s$. The complex Poynting theorem thus \eqref{eq:Poynting_cplx} applies to three sets of currents, fields, and environments: $(\vb{J}_v, \vb{E}_v, \vb{H}_v)$ in vacuum, $(\vb{J}_s, \vb{E}_s, \vb{H}_s)$ in vacuum, and $(\vb{J}_v, \vb{E}_t, \vb{H}_t)$ over the structure, giving
\begin{equation}
    \int_{\partial V_k} \,\dd \vb{\sigma}\cdot(\vb{E}_v\cross\vb{H}^*_v) = i\omega \int_{V_k} \vb{H}_v^* \cdot \vb{H}_v \,\dd V - i\omega  \int_{V_k} \vb{E}_v \cdot \vb{E}_v^* - \int_{V_k} \vb{E}_v \cdot \vb{J}_v^* \,\dd V .
    \label{eq:Poynting_inc}
\end{equation}

\begin{equation} 
    \int_{\partial V_k} \,\dd \vb{\sigma}\cdot(\vb{E}_s\cross\vb{H}^*_s) = i\omega \int_{V_k} \vb{H}_s^* \cdot \vb{H}_s \,\dd V - i\omega \int_{V_k} \vb{E}_s \cdot \vb{E}_s^* \,\dd V - \int_{V_k} \vb{E}_s \cdot \vb{J}_s^* \,\dd V .
    \label{eq:Poynting_sca}
\end{equation}

\begin{equation}
    \int_{\partial V_k} \,\dd \vb{\sigma}\cdot(\vb{E}_t \cross \vb{H}^*_t) 
    = i\omega \int_{V_k} \vb{H}_t^* \cdot \vb{H}_t \,\dd V - i\omega \int_{V_k} (1 + \chi^*\I_s) \vb{E}_t \cdot \vb{E}_t^* \,\dd V - \int_{V_k} \vb{E}_t \cdot \vb{J}_v^* \,\dd V .
    \label{eq:Poynting_tot}
\end{equation}
Subtracting \eqref{eq:Poynting_inc} and \eqref{eq:Poynting_sca} from \eqref{eq:Poynting_tot} gives
\begin{equation} \begin{aligned}
    \Big\{ i \omega \int_{V_k} \vb{H}_v^* \cdot \vb{H}_s \,\dd V - \int_{\partial V_k} \,\dd \vb{\sigma}\cdot(\vb{E}_s \cross \vb{H}_v^*) - i\omega \int_{V_k} \vb{E}_s \cdot \vb{E}_v^* \,\dd V  \Big\} \\
    +\Big\{i \omega \int_{V_k} \vb{H}_s^* \cdot \vb{H}_v \,\dd V - \int_{\partial V_k} \,\dd \vb{\sigma}\cdot(\vb{E}_v \cross \vb{H}_s^*) - i\omega \int_{V_k} \vb{E}_v \cdot \vb{E}_s^* \,\dd V  \Big\} \\
    = \int_{V_k} \vb{E}_s \cdot \vb{J}_v^* \,\dd V - \int_{V_k} \vb{E}_s \cdot \vb{J}_s^* \,\dd V + i\omega \int_V \chi^* \vb{E}_t \cdot \vb{E}_t^* \,\dd V %\numthis 
    \label{eq:Poynting_intermediate}
\end{aligned} \end{equation}
Now, using vector calculus identities along with the Maxwell wave equations $\curl\curl\vb{E}_v - \omega^2 \vb{E}_v = i\omega \vb{J}_v$ and $\curl\curl\vb{E}_s - \omega^2 \vb{E}_s = i\omega \vb{J}_s$, the two curly brackets in \eqref{eq:Poynting_intermediate} can be shown to be equal to $\int_{V_k} \vb{E}_s \cdot \vb{J}_v^* \,\dd V$ and $\int_{V_k} \vb{E}_v \cdot \vb{J}_s^* \,\dd V$ respectively. Finally, the induced current $\vb{J}_s$ can be swapped out by the polarization $\vb{p}$ via $\vb{J}_s = -i\omega \vb{P}$, and the scattered field $\vb{E}_s = \G \vb{P}$, to give
\begin{equation}
    \int_{V_k} \vb{E}_v^* \cdot \vb{P} \,\dd V = \int_V \chi^{-1*} \vb{P}^*\cdot \vb{P} \,\dd V - \int_{V_k} \vb{P}^* \cdot ( \G^\dagger \vb{P}) \,\dd V. 
\end{equation}
This can be written in a compact bra-ket notation
\begin{equation}
    \bra{{E}_v} \I_{V_k} \ket{{P}} = \bra{{P}} (\chi^{-\dagger} - \G^\dagger) \I_{v_k} \ket{{P}},
    \label{eq:QCQP_constraint}
\end{equation}
giving a form of the energy conservation constraints in the text, where \(\vb E_v \to \vb S\). 
From this derivation it is clear that the constraint \eqref{eq:QCQP_constraint} encodes conservation of power during the electromagnetic scattering process for every region $V_k$. 
In the case of many sources (as in this paper), each source defines a different scattering problem, with individual source-polarization pairs $(\vb{S}_j, \vb{P}_j)$ satisfying constraints of the form \eqref{eq:QCQP_constraint}. There are also additional ``cross-constraints'' that capture the fact that the same structured media generates the $\vb{P}_j$ induced in each case:
\begin{equation}
    \bra{ S_j} \ket{ P_k} - \bra{ P_j} \left(\chi^{-\dagger} - \G^\dagger \right) \ket{ P_k} = 0, \quad \forall j,k.
    \label{eq:cross_constraints}
\end{equation}
For a more detailed discussion of these constraints, we refer the reader to~\cite{molesky_mathbbt-operator_2021}.
In the next section, it will be shown that similar constraints can be derived for the Raman problem. 

\section{Derivation of Raman QCQP}
In this section, we show how non-linearly coupled Maxwell's equations can be relaxed to derive optimization objectives and their corresponding constraints to get a QCQP. 
As in the main text, we write Maxwell's equations for Raman scattering as:
\begin{subequations}
\begin{align}
    \nabla \times \nabla \times \vb E_1 - k_1^2 \left( 1+ \chi(\omega_1)\right) \Proj_\chi \vb E_1 &= i \omega_1 \vb{J}_{vac} \\
    \nabla \times \nabla \times \vb{E_2} - k_2^2 \left( 1+ \chi(\omega_2)\right) \Proj_\chi \vb{E_2} &= \omega_2^2 \alpha \Proj_\alpha \vb E_1 = i \omega_2 \vb J_2
    \end{align}
\end{subequations}
where these quantities are defined in the main text. 
\\~\\
\textbf{Design material near a known Raman scatterer.}
The simplest case is a linear design material near a known Raman scatterer. We define $\G_i$ as the operator form of the Green's function of the vacuum Maxwell operator $\M_i$ such that $\M_i \G_i = k_i^2$. By splitting $\G_i$ into the design region and Raman scattering regions, we can write 
\begin{equation}
    \G_i = \begin{bmatrix}
        \G_{i,dd} & \G_{i,dr} \\ 
        \G_{i,rd} & \G_{i, rr}
    \end{bmatrix}, \qquad \Proj_\alpha = \begin{bmatrix} 0 & \I_{r,r} \end{bmatrix}, \qquad \Proj_\chi = \begin{bmatrix} \I_{d,d} & 0 \end{bmatrix},
\end{equation}
where we have defined $\Proj_\chi$ as an operator that projects a full vector (including the design and Raman region) into the design region. 
In bra-ket notation, we keep track of the scattering of the fields:
\begin{subequations}
\begin{align}
    \ket{E_{i,1}} = \dfrac{iZ}{k_1} \G_1 \ket{J_{vac}} \equiv \dfrac{iZ}{k_1} \ket{S_1} \qquad& \text{Known incident field at $\omega_1$} \\
    \ket{E_{s,1}} = \dfrac{iZ}{k_1} \G_{1,dd} \ket{P_1} \qquad& \text{Scattered field in device at } \omega_1 \\
    \ket{J_2}     = -i \omega_2 \alpha \dfrac{iZ}{k_1} (\Proj_\alpha \ket{S} + \G_{1,rd} \ket{P_1})  \qquad& \text{Raman source current at } \omega_2 \\
    \ket{E_{i,2}} = \dfrac{i Z}{k_2} \G_{2,dr} \ket{J_2} \equiv \dfrac{i Z}{k_2} \ket{S_2} \qquad& \text{Raman field produced by Raman material at } \omega_2 \\
    \ket{E_{s,2}} = \dfrac{i Z}{k_2} \G_{2,dd} \ket{P_2} \qquad& \text{Raman field scattered by linear material at } \omega_2
\end{align}
\end{subequations}
where we've defined $\ket{P_1} \equiv \T_1 \G_1 \ket{J_{vac}}$ for the \(\T_i = \left( \chi_i^{-1} - \G_{1,dd} \right)^{-1} \) operator defined in \cite{Molesky_Chao_Jin_Rodriguez_2020}, and $\ket{P_2} \equiv \T_2 \G_2 \ket{J_2}$ is the polarization current in the linear material induced by \(\ket{E_{i,2}}\).
We neglect the effects of $\ket{E_{s,2}}$ further interacting with the Raman scatterer. Our optimization vectors will be the polarization currents $\ket{P_1}, \ket{P_2}$ in the design region.

Let $\chi_1,\chi_2$ be the susceptibilities of the linear material at $\omega_1, \omega_2$ respectively. 
The objectives treated in the text are
\begin{equation} 
f_\text{ext} = \underbrace{-\dfrac12 \Re \bra{J_2}\ket{E_2}}_{\text{extracted power}}, \qquad f_\text{scat} = f_\text{ext} - \underbrace{\bra{P_2} \dfrac{\Im \chi_2}{\abs{\chi_2}^2} \dfrac{Z}{2k_2} \ket{P_2}}_{\text{absorbed power}}.
\end{equation}

Expanding in terms of the optimization variables,
\begin{equation} \begin{aligned} 
 f_\text{ext} = \dfrac12 \Im \dfrac{\omega_2^\dagger}{\omega_1^\dagger \omega_2 }  \Big( & \bra{P_1} \G_{1,rd}^\dagger \alpha^\dagger \G_{2,rd} \ket{P_2} + \bra{S_1} \Proj_\alpha^\dagger \alpha^\dagger \G_{2,rd} \ket{P_2} \Big) \\ 
 + \dfrac12 \Im \dfrac{\omega_2^\dagger}{\abs{\omega_1}^2 } \Big( & \bra{P_1} \G_{1,rd}^\dagger \alpha^\dagger \G_{2,rr} \alpha \G_{1,rd} \ket{P_1} + \bra{S_1} \Proj_\alpha^\dagger \alpha^\dagger \G_{2,rr} \alpha \G_{1,rd} \ket{P_1} \\
 + & \bra{P_1} \G_{1,rd}^\dagger \alpha^\dagger \G_{2,rr} \alpha \Proj_\alpha \ket{S_1} + \bra{S_1} \Proj_\alpha^\dagger \alpha^\dagger \G_{2,rr} \alpha \Proj_\alpha \ket{S_1} \Big).
\end{aligned} \end{equation}
The constraints are derived from identity relations on the \(\T\) operators: \(\Proj_j  = \T_i^\dagger \U_i \Proj_j \) for operators \(\Proj_j\) that project onto any sub-region \(j\) and \(\U_i \equiv (\chi_i^{-\dagger} - \G_{i,dd}^\dagger ) \).
Alternatively, Poynting's theorem may be used as in the previous section. 
Inner products with combinations of \(\ket{S_1}\) and \(\ket{S_2}\) give
\begin{subequations} \begin{align}
    \bra{S_1} p^\dagger \Proj_j \ket{P_1} = \bra{P_1} \U_1 \Proj_j \ket{P_1},     \label{eq:PC_omega1} \\ 
    \dfrac{\omega_2^\dagger}{\omega_1^\dagger} \left( \bra{P_1} \G_{1,rd}^\dagger \alpha^\dagger \G_{2,dr}^\dagger \Proj_j \ket{P_2} + \bra{S_1} \Proj_\alpha^\dagger \alpha^\dagger \G_{2,dr}^\dagger \Proj_j \ket{P_2} \right) = \bra{P_2} \U_2 \Proj_j \ket{P_2}, \label{eq:conservation_2} \\ 
    \bra{S_1} p^\dagger \Proj_j \ket{P_2} = \bra{P_1} \U_1 \Proj_j \ket{P_2}, \\ 
    \dfrac{\omega_2^\dagger}{\omega_1^\dagger} \left( \bra{P_1} \G_{1,rd}^\dagger \alpha^\dagger \G_{2,dr}^\dagger \Proj_j \ket{P_1} + \bra{S_1} \Proj_\alpha^\dagger \alpha^\dagger \G_{2,dr}^\dagger \Proj_j \ket{P_1} \right) = \bra{P_2} \U_2 \Proj_j \ket{P_1}.
\end{align} \end{subequations}
The dual problem is solved using an interior point method, requiring an initial feasible point. To find it, we enforce the passivity constraint \( \dfrac{\Im \chi}{\lvert \chi \rvert^2} \bra{P_2}\ket{P_2} \leq \Asym \bra{P_2} \G_{2,dr} \ket{J_2} \), which is trivially satisfied if~\eqref{eq:conservation_2} is satisfied. Noting that 
\begin{equation*} 
\Asym \bra{P_2} \G_{2,dr} \ket{J_2} \leq \sqrt{\braket{P_2}} \sqrt{\bra{J_2} \G_{2,dr}^\dagger \G_{2,dr} \ket{J_2}}
\end{equation*} 
and expanding in terms of optimization variables, the convex inequality constraint becomes
\begin{equation}\label{eq:cc}
    -\left( \dfrac{\Im \chi}{\lvert \chi \rvert^2} \right)^2 \braket{P_2} + \dfrac{\abs{\omega_2}^2}{\abs{\omega_1}^2 } \left( \bra{S_1} \Proj_\alpha^\dagger + \bra{P_1} \G_{1,rd}^\dagger \right) \alpha^\dagger \G_{2,dr}^\dagger \G_{2,dr} \alpha \Big( \Proj_\alpha \ket{S_1} + \G_{1,rd} \ket{P_1} \Big) \geq 0.
\end{equation} 
To find an initial feasible point, where the matrix \(Z^{TT}\) is positive-definite and invertible, we note that the positive semi-definiteness of \(\Im \G_{i,dd}\) and the assumption that \(\Im \chi > 0\) guarantees that for some small Lagrange multiplier corresponding to Eq.~\ref{eq:cc}, there exists some sufficiently large Lagrange multiplier corresponding to Eq.~\ref{eq:PC_omega1} that makes \(Z^{TT}\) positive definite. 
In this case, \(P_\text{opt} = [\ket{P_1}, \ket{P_2}]^T\).
\\~\\
\textbf{Raman Active Design Material.}
We take $ \Proj_\alpha = \Proj_\chi $ such that the Raman scatterer is only present where material is present. To ensure bounds are structure independent, we relax $\ket{J_2}$ to an optimization degree of freedom. The scattered power objective is:
\begin{equation}
    f_{\text{scat}} = \dfrac{Z}{2 k_2} \left[ \Im \left( \bra{J_2} \G_0^{(2)} \ket{J_2} + \bra{J_2} \G_0^{(2)} \ket{P_2} \right) - \bra{P_2} \dfrac{\Im \chi_2}{\lvert \chi_2 \rvert^2} \ket{P_2} \right].
\end{equation}
The constraints are derived from the same identities as above. Inner products with incident fields and \(\ket{J_2}\) give
\begin{subequations} \begin{align}
    \bra{S_1} \Proj_j \ket{P_1} &= \bra{P_1} \U_1 \Proj_j \ket{P_1}, \\ 
    \bra{J_2} \G_2^{\dagger} \Proj_j \ket{P_2} &= \bra{P_2} \U_2 \Proj_j \ket{P_2}, \\ 
    \bra{S_1} \Proj_j \ket{P_2} &= \bra{P_1} \U_1 \Proj_j \ket{P_2}, \\ 
    \bra{S_1} \Proj_j \ket{J_2} &= \bra{P_1} \U_1 \Proj_j \ket{J_2}, \\
    \bra{J_2} \G_2^{\dagger} \Proj_j \ket{P_1} &= \bra{P_2} \U_2 \Proj_j \ket{P_1}.
\end{align} \end{subequations}
By taking an inner product of \(\ket{J_2}\) with itself, we find the additional constraint \(\bra{J_2}\Proj_j\ket{J_2} d\vb r = \abs{\omega_2 \alpha}^2 \bra{\vb E_1} \Proj_\alpha^\dagger \Proj_j \Proj_\alpha \ket{\vb E_1})\). Because \(\Proj_\alpha = \Proj_\chi\) and \(\ket{P_1} = \frac{k_1}{iZ} \chi \Proj_\chi \ket{E_1}\), this constraint can be written
\begin{equation} \begin{aligned}
    \bra{J_2} \Proj_j \ket{J_2} &=  \abs{\dfrac{\omega_2 \alpha }{\omega_1 \chi_1}}^2 \bra{P_1} \Proj_j \ket{P_1}.
\end{aligned} \end{equation}
In this case, \(P_\text{opt} = [\ket{P_1}, \ket{P_2}, \ket{J_2}]^T\). 

\section{Bounding Decoupled Raman-Enhancement Problem}
\noindent
\textbf{Focusing.} The field-focusing optimization problem takes the form
\begin{equation} \label{eq:focusing} \begin{aligned}
    \underset{\ket{P_1}}{\text{max}} \quad & \dfrac{Z^2}{\abs{k_1}^2} \left[ \bra{S_1} \Proj_\alpha^\dagger \alpha^\dagger \alpha \Proj_\alpha \ket{S_1} + 2\Re \bra{P_1} \G_{1,rd}^\dagger \Proj_\alpha^\dagger \alpha^\dagger \alpha \Proj_\alpha \ket{S_1} + \bra{P_1} \G_{1,rd}^\dagger \Proj_\alpha^\dagger \alpha^\dagger \alpha \Proj_\alpha \G_{1,rd} \ket{P_1} \right] \\ 
    \text{s.t.} \quad & \bra{S_1} \Proj_j \ket{P_1} - \bra{P_1} \U_1 \Proj_j \ket{P_1} = 0
\end{aligned} \end{equation}
A bound \(b_\text{focus}\) can be computed via the Lagrange dual as above where \(\psi_\text{opt} = \vb P_1 \). 
The bound on the amplitude \(\braket{J_2}\) of the Raman-active dipole is therefore \(M_f \equiv \omega_2^2 b_\text{focus} \).  
\\~\\
\textbf{Scattered power.} The scattering optimization problem takes the form
\begin{equation} \begin{aligned}
    \underset{\ket{P_2}}{\text{max}} \quad & \dfrac{Z}{2 k_2} \Im \bra{S_2} \ket{P_2} - \dfrac{Z}{2k_2} \bra{P_2} \dfrac{\Im \chi_2}{\abs{\chi_2}^2} \ket{P_2} \\
    \text{s.t.} \quad & \bra{S_2} \Proj_j \ket{P_2} - \bra{P_2} \U_2 \Proj_j \ket{P_2} = 0
\end{aligned} \end{equation}
where in the Raman molecule case, the incident field from the Raman scatterer \(\ket{S_2}\) is treated as a unit dipole field. A bound \(M_l\) can be computed using the methods above where \(\ket{\psi_\text{opt}} = \ket{P_2} \). \\~\\
The final decoupled bound on the Raman enhancement problem is \(M_f M_l\).

\section{Comparison to Existing Bounds}
\begin{figure}
    \centering
    \includegraphics[width=\linewidth]{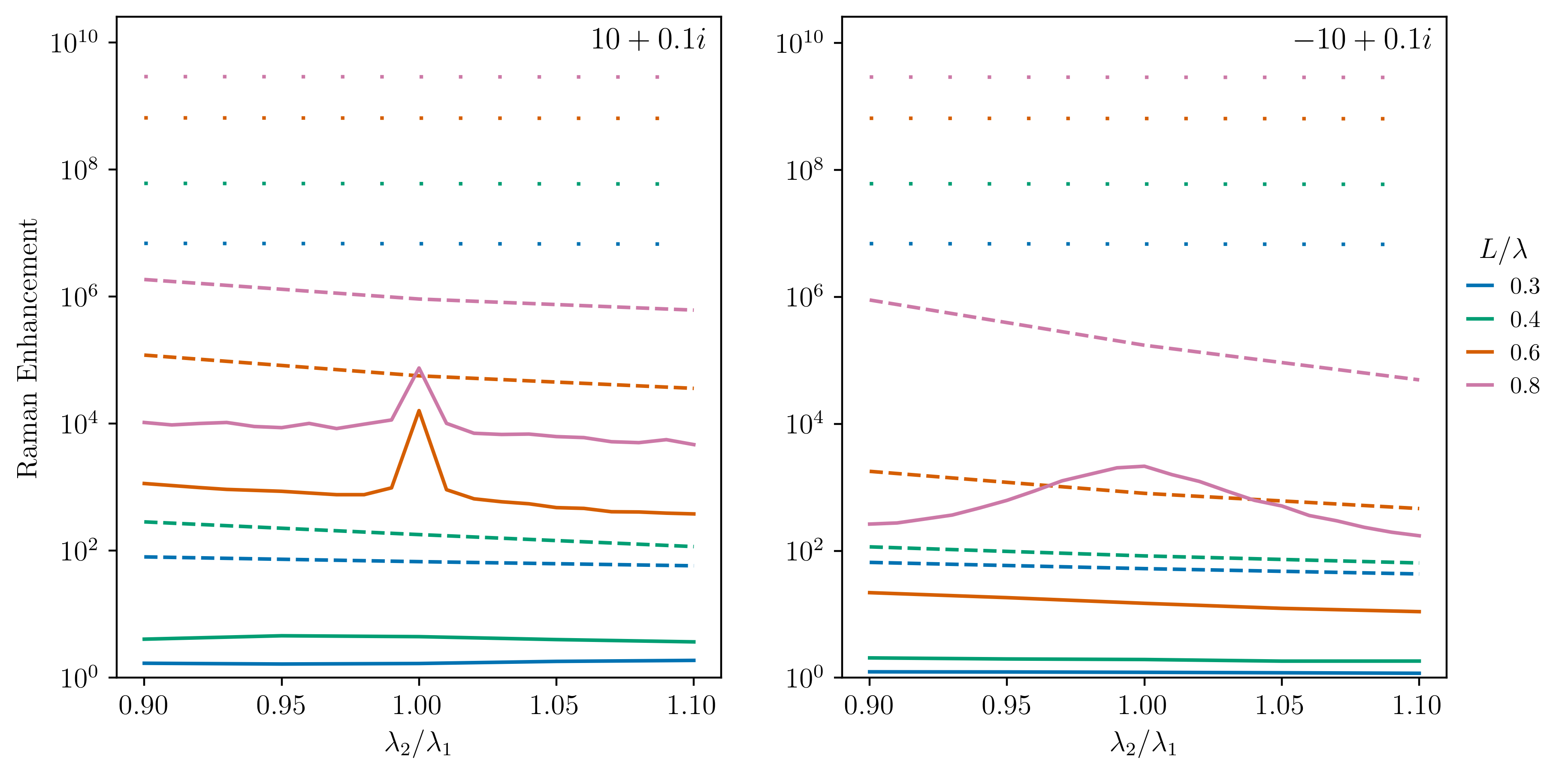}
    \caption{Bounds on the Raman molecule problem as described in the main text, replacing \(f_\text{scat} \to f_\text{ext}\) for fair comparison to existing work. Limits are computed with our coupled method (solid), our decoupled method (dashed) and methods in~\cite{Michon_Benzaouia_Yao_Miller_Johnson_2019} (dotted) for a variety of design sizes \(L/\lambda = \{0.3, 0.4, 0.6, 0.8\}\). These results make clear the significant improvement in these bounds achieved by the methods in this article.}
    \label{fig:active_comparison}
\end{figure}

For a fair comparison of our bounds with existing methods, we calculate analogous bounds on \(f_{ext}\) for the Raman particles problem seen in the main text with the methods in~\cite{Michon_Benzaouia_Yao_Miller_Johnson_2019}. 
Although these bounds were found to be tight in specific cases, our method exhibited many orders of magnitude improvement for the case studied in this paper. A direct comparison is shown in Figure~\ref{fig:active_comparison}.

\section{Inverse Design Details}
The problem of inverse design is that of solving Eq.~(2) in the main text for a given objective function \(f_0(\vb E_1, \vb E_2)\). The steps to solve this problem are outlined below:
\begin{enumerate}
    \item A Maxwell FDFD solver was written in JAX~\cite{jax2018github} to solve \(\left( \nabla \times \nabla \times - k^2 (1+\chi_d) \Proj_\chi \right) \vb E \equiv \M \vb E = i\omega \vb J\) for a field \(\vb E\). Using JAX's automatic differentiation tools, derivatives of any function of \(\vb E \) with respect to changes in \(\Proj_\chi\) can be computed automatically. 
    \item In the spirit of other topology optimization routines, \(\Proj_\chi\) is relaxed to a continuous parameter. That is, \(\chi \Proj_\chi \) may take values in \(\chi_d \I + (1-\chi_d) \rho(\vb r) \) for \(\rho(\vb r) \in [0,1]\), anywhere. \(\rho\) will be the optimization degree of freedom. 
    \item \(\Proj_\chi\) is initialized to be random and a known initial current \(\vb J_{\text{vac}}\) is simulated using a linear Maxwell FDFD solve, thereby solving Eq. 1a in the main text. 
    \item Using known \(\alpha\) and \(\Proj_\alpha\) (which for the inverse design problem is known for both the molecule and medium case), the Raman current \(\vb J_2\) is computed.
    \item Another Maxwell FDFD solve is performed, solving Eq. 1b in the main text. 
    \item Knowing the fields \(\vb E_1\) and \(\vb E_2\), any objective of the fields can be calculated. In the case of this paper, it is the objective given in Eq. 2 in the main text. 
    \item Because the FDFD solver is written in JAX and can be automatically differentiated, d(objective)/d\(\Proj_\chi\) may be calculated using automatic differentiation methods. This is analogous to an adjoint optimization, but saves lengthy manual calculation of the adjoint. Adjoint solves of the Maxwell operator are still required to compute this derivative. 
    \item Having defined an objective function over \(\rho\), JAXopt~\cite{jaxopt_implicit_diff} can be utilized to perform projected gradient descent to maximize the objective.
    \item In this paper, structures are not binarized for best comparison to bounds, which represent limits on unbinarized devices. 
\end{enumerate}

\bibliographystyle{plain}
\bibliography{supp}